\documentclass[a4paper,nohyper,11pt]{JHEP3}

\usepackage{amstext,amsfonts,amsmath,amssymb} 
\usepackage[all, knot]{xy}

\providecommand{\bysame}{\leavevmode\hbox to3em{\hrulefill}\thinspace}
\providecommand{\MR}{\relax\ifhmode\unskip\space\fi MR }

\providecommand{\href}[2]{#2}

\newfont{\gl}{eufm10 scaled \magstep1} 

\newcommand{\tr}{\hbox{tr}}
\newcommand{\dd}{\hbox{d}}
\newcommand{\Ad}{\hbox{Ad}}
\newcommand{\ad}{\hbox{ad}}

\newcommand{\gp}{{g_{_+}}} 
\newcommand{\gm}{{g_{_-}}} 
\newcommand{\gpinv}{{g_{_{+}}^{-1}}}
\newcommand{\gminv}{{g_{_{-}}^{-1}}}

\title{Dual branes in topological sigma models over Lie
groups. BF-theory and non-factorizable Lie bialgebras}

\author{Iv\'an Calvo and Fernando Falceto\\ Departamento de
F\'{\i}sica Te\'orica, Universidad de Zaragoza\\E-50009 Zaragoza,
Spain\\ E-mail: \email{icalvo@unizar.es,
falceto@unizar.es}\thanks{Work supported by MEC (Spain), grant
FPA2003-02948. I. C. is supported by MEC (Spain), grant FPU.}}

\abstract{We continue the study of the Poisson-Sigma model over
Poisson-Lie groups. Firstly, we solve the models with targets $G$ and
$G^*$ (the dual group of the Poisson-Lie group $G$) corresponding to a
triangular $r$-matrix and show that the model over $G^*$ is always
equivalent to BF-theory. Then, given an arbitrary $r$-matrix, we
address the problem of finding D-branes preserving the duality between
the models. We identify a broad class of dual branes which are
subgroups of $G$ and $G^*$, but not necessarily Poisson-Lie
subgroups. In particular, they are not coisotropic submanifolds in the
general case and what is more, we show that by means of duality
transformations one can go from coisotropic to non-coisotropic
branes. This fact makes clear that non-coisotropic branes are natural
boundary conditions for the Poisson-Sigma model.}

\begin{document}

\section{Introduction}

The Poisson-Sigma model is a two-dimensional topological sigma model
defined on a surface $\Sigma$ and whose target is a Poisson manifold
$M$. The fields of the model are given by a bundle map
$(X,\psi):T\Sigma \rightarrow T^*M$ where $X:\Sigma \rightarrow M$ is
the base map and $\psi\in \Gamma(T^*\Sigma\otimes X^*T^*M)$.

The model has given much insight in a variety of topics on Poisson
Geometry, since it encodes in a very natural way the geometrical
properties of the target. Examples are the relation of the
perturbative quantization of the model to Kontsevich's formula for
deformation quantization (\cite{Kon},\cite{CatFel99},\cite{CatFel03},
\cite{CalFal051}) as well as the connection to symplectic groupoids
(\cite{CatFel00}).

A Poisson-Lie group is a Lie group equipped with a Poisson structure
which is compatible with the product on the group. The
particularization of the model to this case, which we shall call {\bf
Poisson-Lie sigma model}, is an interesting topic on its own. Each
Poisson-Lie group $G$ has a dual group $G^*$ which is also Poisson-Lie
and one expects this duality to show up in the models.

Every Poisson-Lie structure on a complex simple Lie group $G$ is given
by an $r$-matrix either factorizable or triangular (see the following
sections). The Poisson-Sigma model in the factorizable case was
studied in \cite{CalFalGar}. It was discovered a bulk-boundary duality
relating the reduced phase space of the models with targets $G$ and
$G^*$ when $\Sigma={\mathbb R}\times [0,\pi]$ and the base map
$g:\Sigma\rightarrow M$ is free at the boundary of
$\Sigma$. Triangular $r$-matrices were not treated therein due to the
absence of an efficient realization of the double group in this
case. Now, we can describe explicitly the double group for any
Poisson-Lie structure on a complex, simple and simply connected Lie
group. We make use of such description for studying in the Lagrangian
and Hamiltonian formalisms the Poisson-Sigma model with a triangular
Poisson-Lie group as target. We show that the duality found in
\cite{CalFalGar} shows up also in the triangular case.

 In \cite{AleSchStr},\cite{FalGaw02} it was shown that, in the
factorizable case, the Poisson-Sigma model with $G^*$ as target is
locally equivalent to the $G/G$ WZW model. We prove in the present
paper that when $r$ is triangular, the Poisson-Sigma model with target
$G^*$ is always equivalent to BF-theory.

The task of identifying the most general branes (that is, submanifolds
to which $X\vert_{\partial\Sigma}$ can be restricted) which are
admissible for the Poisson-Sigma model was started in \cite{CatFel03}
and completed in \cite{CalFal04}. The main objective of the present
paper is to apply the machinery of \cite{CalFal04} to the Poisson-Lie
sigma models over $G$ and $G^*$ and find branes which preserve and
generalize the bulk-boundary duality found in \cite{CalFalGar} in the
particular case in which the brane was the whole target group. This
duality transformation gave a one-to-one map between the moduli spaces
of solutions of the models over $G$ and $G^*$. The dualizable branes
should be compatible with the group and Poisson-Lie structures. We
identify a family of dualizable branes which are $r$-invariant
subgroups, but not necessarily coisotropic subgroups. More
interestingly, we show that the dual brane of a coisotropic brane can
be non-coisotropic. This explains why focusing on pairs of coisotropic
branes (\cite{BonZab}) leads to different moduli spaces of solutions.

\vskip 0.2 cm

The paper is organized as follows:

In Section 2 we give a brief survey on Poisson-Lie groups and Lie
bialgebras and in Section 3 we describe the double of a Lie bialgebra
(and the corresponding double group) adopting an approach that allows
to treat both the factorizable and triangular case and understand
their differences. Section 4 presents the formulation of the
Poisson-Sigma model on surfaces with boundary and a summary of the
results of \cite{CalFal04} about admissible branes. In Section 5 we
recall the results of \cite{CalFalGar} on the factorizable case and
solve in detail the models corresponding to triangular
$r$-matrices. We show that the bulk-boundary duality pointed out in
\cite{CalFalGar} for free boundary conditions is still present in the
triangular case. We also prove in the Lagrangian formalism that if $r$
is triangular, the Poisson-Sigma model with target $G^*$ is equivalent
to BF-theory. Section 6 deals with the problem of dual branes.

\section{Poisson-Lie groups} \label{introPL}

We present here some basic results on Poisson-Lie groups and Lie
bialgebras. See \cite{Sem85},\cite{LuWei} and the introductory notes
\cite{Kos} for details.

A {\bf Poisson manifold} $M$ is a differentiable manifold endowed with
a bilinear bracket $\{\cdot,\cdot\}:C^\infty(M)\times C^\infty(M)\rightarrow
C^\infty(M)$ verifying:

$(i)$ $\{f_1,f_2\}=-\{f_2,f_1\}$

$(ii)$ $\{f_1,\{f_2,f_3\}\}+\{f_2,\{f_3,f_1\}\}+\{f_3,\{f_1,f_2\}\}=0$
\hfill (Jacobi identity)

$(iii)$ $\{f_1,f_2f_3\}=f_2\{f_1,f_3\}+\{f_1,f_2\}f_3$ \hfill (Leibniz rule)

\noindent for any $f_1,f_2,f_3 \in C^\infty(M)$.

\vskip 0.2 cm

The Poisson bracket $\{\cdot,\cdot\}$ determines uniquely a bivector field $\Pi$
such that:
$$\{f,g\}(p)=\iota(\Pi_p) (df\wedge dg)_{p},\ p \in M$$

Taking local coordinates $X^i$ on $M$, $\Pi^{ij}(X)=\{X^i,X^j\}$. The
Jacobi identity for the Poisson bracket reads in terms of $\Pi^{ij}$:
$$
\Pi^{ij}
\partial_{i}\Pi^{kl}
+
\Pi^{ik}
\partial_{i}\Pi^{lj}
+
\Pi^{il}
\partial_{i}\Pi^{jk}
=0
$$
where summation over repeated indices is understood. 

\vspace{0.3 cm}

For every $p\in M$ define $\Pi^\sharp_p:T_p^*M\rightarrow T_pM$ by
$$\beta(\Pi^\sharp_p(\alpha))=\iota(\Pi_p) (\alpha \wedge \beta),\ 
\alpha,\beta \in T^*_pM$$

Due to the Jacobi identity the image of $\Pi^{\sharp}$,
$${\rm Im}(\Pi^{\sharp}):=\bigcup_{p\in M}{\rm Im}(\Pi^{\sharp}_{p})$$
is a completely integrable general differential distribution and $M$
admits a generalized foliation (\cite{Vai94}). The Poisson structure
restricted to a leaf is non-degenerate and hence defines a symplectic
structure on the leaf. That is why one often speaks about the
symplectic foliation and the {\bf symplectic leaves} of $M$.

Let $N$ be a submanifold of the Poisson manifold $M$ 
with $i:N \hookrightarrow M$ the inclusion map, 
and denote by $T_pN^0\subset T^*_pM$ the annihilator of $T_pN,p\in N$.
If 
\begin{equation}\label{PoiDir}
\Pi^\sharp_p(T_pN^0)\cap T_pN=\{0\},\quad\forall p\in N
\end{equation}
then for any $p\in N$
there exists a map  $\hat\Pi^\sharp_p$ that makes the following diagram
\begin{equation}
\begin{aligned}
\xymatrix{
  T_p^*N\ar[rr]^{\hat\Pi^\sharp_p}
    &&T_pN\ar[dd]^{{i_{*p}}}\\
  \\
  \Pi_p^{\sharp-1}(T_pN)\ar[rr]_{\Pi_p^\sharp}\ar[uu]^{i^*_{p}}&&T_pM
}
\end{aligned}
\end{equation}
commutative. If the maps $\hat\Pi^\sharp_p$ define a smooth bundle map
$\hat\Pi^\sharp:T^*N\rightarrow TN$ the latter gives a Poisson
structure on $N$ (the Dirac bracket) and $N$ is called {\bf
Poisson-Dirac submanifold} (\cite{CraFer}). On the other hand, $N$ is
said {\bf coisotropic} if $\Pi^\sharp_p(T_pN^0)\subset T_pN,\ \forall
p\in N$. A submanifold $N$ which is both Poisson-Dirac and coisotropic
satisfies $\Pi^\sharp_p(T_pN^0)=0,\ \forall p \in N$ and is said a
{\bf Poisson submanifold}. Equivalently, $N$ is a Poisson submanifold
if the inclusion $i$ is a Poisson map.

\vskip 0.4 cm

Consider a Lie group $G$ equipped with a Poisson structure
$\{\cdot,\cdot\}_{G}$. 
It is natural to demand the Poisson structure to be compatible with
the product.
$G$ is said to be a {\bf Poisson-Lie group} if the product on the
group is a Poisson map, i.e. if
\begin{eqnarray}\label{poissoncomp}
\{f,h\}_{G}(g_{1}g_{2})=\{f(\cdot g_{2}),h(\cdot g_{2})\}_{G}(g_{1}) + \{f(g_{1}\cdot),h(g_{1}\cdot)\}_{G}(g_{2})
\end{eqnarray}
for $f,h \in C^{\infty}(G)$. It is evident from (\ref{poissoncomp})
that a Poisson-Lie structure always vanishes at the unit $e$ of
$G$. Therefore, the linearization of the Poisson structure at $e$
provides a Lie algebra structure on ${\mathfrak g}^*=T^*_{e}(G)$ by
the formula
\begin{eqnarray} \label{Liedual}
[\dd f_1(e),\dd f_2(e)]_{{\mathfrak g}^*}=\dd \{f_1,f_2\}_G(e),\
f_1,f_2 \in C^\infty(G)
\end{eqnarray}

The Poisson-Lie structure of $G$ yields a compatibility condition
between the Lie brackets on ${\mathfrak g}$ and ${\mathfrak g}^*$,
namely
\begin{eqnarray}\label{comp} 
\left<[\xi,\eta]_{{\mathfrak g}^*},[X,Y]\right>&+&\left<\ad_Y^*\eta,
\ad_\xi^*X\right>-\left<\ad_Y^*\xi,\ad_\eta^*X\right>-\cr
&-&\left<\ad_X^*\eta,\ad_\xi^*Y\right>+\left<\ad_X^*\xi,\ad_\eta^*Y\right>=0.
\end{eqnarray}
for $X,Y \in {\mathfrak g},\ \xi,\eta \in {\mathfrak g}^*$ and
$\left<\cdot,\cdot\right>$ the natural pairing between elements of a vector
space and its dual.\footnote{$\ad ^*$ denotes the coadjoint
representation of a Lie algebra on its dual vector space. Hence, $\xi
\in {\mathfrak g}^* \mapsto \ad ^*_\xi$ is the coadjoint
representation of $({\mathfrak g}^*,[\cdot,\cdot]_{{\mathfrak g}^*})$
on ${\mathfrak g}$.}.

The compatibility condition (\ref{comp}) between the Lie brackets on
${\mathfrak g}$ and ${\mathfrak g}^*$ defines a {\bf Lie bialgebra}
structure for ${\mathfrak g}$ (or, by symmetry, for ${\mathfrak
g}^*$).

Now take ${\mathfrak g}\oplus{\mathfrak g}^*$ with the natural scalar product
\begin{eqnarray} \label{scalarproduct}
&&(X+\xi \vert Y+\eta)=\left<\eta,X\right>+\left<\xi,Y\right>,\ X,Y \in
{\mathfrak g},\ \xi,\eta \in {\mathfrak g}^*
\end{eqnarray}
There exists a unique Lie algebra structure on ${\mathfrak
g}\oplus{\mathfrak g}^*$ such that ${\mathfrak g}$ and ${\mathfrak
g}^*$ are Lie subalgebras and that (\ref{scalarproduct}) is invariant:
\begin{eqnarray} \label{Liesuma}
&&[X+\xi,Y+\eta]=[X,Y]+[\xi,\eta]_{{\mathfrak g}^*}-\ad^{*}_{X}\eta 
+ \ad^{*}_{Y}\xi + \ad^{*}_{\eta}X - \ad^{*}_{\xi}Y
\end{eqnarray}

The vector space ${\mathfrak g}\oplus {\mathfrak g}^*$ with the Lie bracket
(\ref{Liesuma}) is called the double of ${\mathfrak g}$ and is denoted
by ${\mathfrak g}\bowtie{\mathfrak g}^*$ or ${\mathfrak d}$.

If $G$ is connected and simply connected, (\ref{comp}) is enough to
integrate $[\cdot,\cdot]_{{\mathfrak g}^*}$ to a Poisson structure on
$G$ that makes it Poisson-Lie and the Poisson structure is
unique. Hence, there is a one-to-one correspondence between
Poisson-Lie structures on $G$ and Lie bialgebra structures on
${\mathfrak g}$.  The symmetry between ${\mathfrak g}$ and ${\mathfrak
g}^*$ in (\ref{comp}) implies that one has also a Poisson-Lie group
$G^*$ with Lie algebra $({\mathfrak g}^*,[\cdot,\cdot]_{{\mathfrak
g}^*})$ and a Poisson structure $\{\cdot ,\cdot \}_{G^*}$ whose
linearization at $e$ gives the Lie bracket of ${\mathfrak g}$. $G^*$
is the {\bf dual Poisson-Lie group} of $G$. The connected and simply
connected Lie group with Lie algebra ${\mathfrak g}\bowtie{\mathfrak
g}^*$ is known as the {\bf double group} of $G$ and denoted by $D$.

$G$ and $G^*$ are subgroups of $D$ and there exists a neighborhood
 ${D_0}$ of the identity of $D$ such that every element $\nu \in
 {D_0}$ can be written as $\nu = ug = {\tilde g}{\tilde u},\
 g,{\tilde g}\in G, \ u,{\tilde u} \in G^{*}$ and both factorizations
 are unique (notice that $G_0 := G \cap G^* \subset D$ is a
 discrete subgroup). These factorizations define a local left action
 of $G^{*}$ on $G$ and a local right action of $G$ on $G^{*}$ by
\begin{eqnarray}
&&{}^{u}g = {\tilde g} \cr
&&u^{g} = {\tilde u}
\end{eqnarray}

Starting with the element $gu \in D$ we can define in an
analogous way a left action of $G$ on $G^{*}$ and a right action of
$G^{*}$ on $G$. These are known as {\bf dressing transformations} or
{\bf dressing actions}. The symplectic leaves of $G$
(resp. $G^{*}$) are the connected components of the orbits of the
right or left dressing action of $G^{*}$ (resp. $G$).

There is a natural Poisson structure on $D$ which will be important
for us since it will show up in the analysis of the reduced phase
space of the Poisson-Lie sigma models. Its main symplectic leaf (which
contains a neighbourhood of the unit of $D$) is $D_0=GG^*\cap
G^*G$. We write its inverse in $D_0$, which is a symplectic form
defined at a point $ug = {\tilde g}{\tilde u} \in D_0$ as:
\begin{equation} \label{HeisenbergDouble}
\Omega (ug) = \left<\dd\tilde{g}\tilde{g}^{-1}\stackrel{\wedge}{,}\dd{u}{u}^{-1}\right> +
\left<g^{-1}\dd g\stackrel{\wedge}{,}\tilde{u}^{-1}\dd\tilde{u}\right>
\end{equation}
where $\left<\cdot,\cdot\right>$ acts on the values of the
Maurer-Cartan one-forms. $D$ endowed with the Poisson structure
yielding $\Omega$ is known as {\bf Heisenberg double}
(\cite{FalGaw91},\cite{AleMal}).

A {\bf Poisson-Lie subgroup} $H \subset G$ is a subgroup which is
Poisson-Lie and such that the inclusion $i:H \hookrightarrow G$ is a
Poisson map. In particular $H$ is a coisotropic submanifold of
$G$. Let us call ${\mathfrak h}\subset {\mathfrak g}$ the Lie algebra
of $H$ and ${\mathfrak h}^0\subset {\mathfrak g}^*$ its
annihilator. $H$ is a Poisson-Lie subgroup if and only if ${\mathfrak
h}^0$ is an ideal of ${\mathfrak g}^*$, i.e. $[\xi,\eta]_{{\mathfrak
g}^*}\in {\mathfrak h}^0,\ \forall \xi\in {\mathfrak
g}^*,\forall\eta\in {\mathfrak h}^0$. This property permits to
restrict the bialgebra structure to ${\mathfrak h}$, which is then
said a {\bf Lie subbialgebra} of ${\mathfrak g}$. The Poisson-Lie
group $H$ associated to ${\mathfrak h}$ is a subgroup of $G$. However,
in general there is no natural way to realize the dual Poisson-Lie
group $H^*$ as a subgroup of $G^*$\footnote{In reference \cite{BonZab}
$H$ and $H^0$ (the subgroup corresponding to ${\mathfrak h}^0$) were
proposed as a pair of dual branes for the Poisson-Lie sigma models
over $G$ and $G^*$. We shall see later on in this paper that this is
not the right approach. One must take $H^*$ as the dual brane of
$H$.}.

\subsection{Poisson-Lie structures on simple Lie groups}

Let us take G a complex, simple, connected and simply connected Lie
group and give the above construction explicitly. The (essentially
unique) nondegenerate, invariant, bilinear form ${\rm tr}(\ )$ on
${\mathfrak g}$ establishes an isomorphism between ${\mathfrak g}$ and
${\mathfrak g}^*$. The Poisson structure $\Pi$ contracted with the
right-invariant forms on $G$, $\theta_R (X)=\tr(dgg^{-1}X),\ X \in
{\mathfrak g}$, will be denoted
\begin{equation} \label{PoissonContracted}
P_{g}(X,Y)=\iota(\Pi_{g})\theta_R(X)\wedge\theta_R(Y)
\end{equation}

For a general Poisson-Lie structure on G (\cite{LuWei}),
\begin{eqnarray} \label{PoissonLie}
P_{g}^{r}(X,Y)={\frac{1}{2}}\ \tr(XrY-X\Ad_{g}r\Ad_{g}^{-1}Y)
\end{eqnarray}
where $r:{\mathfrak g}\rightarrow {\mathfrak g}$ is an antisymmetric
endomorphism such that
\begin{eqnarray} \label{YangBaxter}
r[rX,Y]+r[X,rY]-[rX,rY]=\alpha[X,Y],\ \alpha \in \mathbb{C}
\end{eqnarray}
which is sometimes called {\bf modified Yang-Baxter identity}. Such an
operator is what we shall understand by an {\bf
$r$-matrix}\footnote{In the literature what we call $r$ is often
denoted by $R$, keeping $r$ for elements of ${\mathfrak
g}\otimes{\mathfrak g}$.}.

It is possible to show that $\Ad_{g_{0}}r = r\Ad_{g_{0}},\ g_{0} \in
G_{0}$.

Using $r$ we can define a second Lie bracket on ${\mathfrak g}$,
\begin{eqnarray} \label{rproduct}
[X,Y]_{r}={\frac{1}{2}}([X,rY]+[rX,Y])
\end{eqnarray}
which is nothing but the linearization of (\ref{PoissonLie}) at the unit
of $G$. Denoting by ${\mathfrak g}_r$ the vector space ${\mathfrak g}$
equipped with the Lie bracket $[\cdot,\cdot]_{r}$, we have that ${\mathfrak
g}_r$ is isomorphic to $({\mathfrak g}^{*},[\cdot,\cdot]_{{\mathfrak g}^*})$.

In fact, every Lie bialgebra structure on a simple Lie algebra is
given by an $r$-matrix as defined above. The pair $({\mathfrak g},r)$
is said a {\bf factorizable} (resp. {\bf non-factorizable} or {\bf
triangular}) Lie bialgebra if $\alpha\neq 0$ (resp. $\alpha=0$). The
corresponding Poisson-Lie groups will be said either factorizable or
triangular accordingly.

Using the isomorphism given by ${\rm tr}(\ )$ it is easy to show that
${\mathfrak g}\bowtie{\mathfrak g}^*\cong({\mathfrak
g}\oplus{\mathfrak g},[\cdot,\cdot]_{\mathfrak d})$ as Lie algebras, where
\begin{subequations}
\begin{align}
&[(X,Y),(X',Y')]_{\mathfrak d} = \notag \\
&([X,X']+{\frac{1}{2}}([X,rY']+[rY,X']+r[Y',X]+ \notag
r[X',Y]),[X,Y']+[Y,X']+[Y,Y']_{r})
\end{align}
\end{subequations}

\section{The double of a Lie bialgebra}

The study of the Poisson-Sigma model on factorizable Poisson-Lie
groups carried out in \cite{CalFalGar} made use of a concrete
realization of the double group. It turns out that the double
${\mathfrak d}$ has a very different aspect for $\alpha \neq 0$ and
$\alpha=0$ and the analysis therein did not cover the triangular
case. In this section we rederive an approach 
(appeared already in \cite{Sto})
that allows us to understand the cases $\alpha \neq 0$ and $\alpha =
0$ in a unified way.

Consider the Lie algebra ${\mathfrak G}={\mathfrak g}[[\varepsilon]]$ of
polynomials on a variable $\varepsilon$ with coefficients in
${\mathfrak g}$ (always a simple complex Lie algebra in this paper)
where
\begin{eqnarray} \label{loopalgebra}
[\sum_{m=0}^{M} X_m \varepsilon^m,\sum_{n=0}^{N} X'_n
\varepsilon^n]=\sum_{m=0}^{M}\sum_{n=0}^{N}\varepsilon^{m+n}[X_m,X'_n]
\end{eqnarray}

${\mathfrak G}_{\alpha}=(\varepsilon^2 - \alpha){\mathfrak G}$ is an
ideal of ${\mathfrak G}$, so ${\mathfrak G}/{\mathfrak G}_{\alpha}$
inherits a Lie algebra structure from ${\mathfrak G}$. In practice, it
is more useful to think of ${\mathfrak G}/{\mathfrak G}_{\alpha}$ as
the set $\{X+Y\varepsilon,\ X,Y \in {\mathfrak g} \mid {\varepsilon}^2
= \alpha\}$. Then, the Lie bracket of two elements of ${\mathfrak
G}/{\mathfrak G}_{\alpha}$ can be written as
\begin{eqnarray} \label{quotient}
[X+Y\varepsilon,X'+Y'\varepsilon]= [X,X']+\alpha [Y,Y'] +
([X,Y']+[Y,X'])\varepsilon
\end{eqnarray}

There exists an isomorphism of Lie algebras between $({\mathfrak
g}\oplus{\mathfrak g},[\cdot,\cdot]_{\mathfrak d})$ and ${\mathfrak
G}/{\mathfrak G}_{\alpha}$ given by $(X,Y)\mapsto X +
{\frac{1}{2}}rY+{\frac{1}{2}}Y\varepsilon$ and, consequently,
${\mathfrak d}\cong {\mathfrak G}/{\mathfrak
G}_{\alpha}$. Furthermore,
$${\mathfrak g}\cong \{X \mid X \in {\mathfrak g}\}, \ {\mathfrak g}_r
\cong \{rX + X\varepsilon \mid X\in {\mathfrak g}\} \subset {\mathfrak
G}/{\mathfrak G}_{\alpha}$$

\subsection{Factorizable Lie bialgebras}

It is clear that if $\alpha \neq 0$, the subalgebras
$\{(1+{\varepsilon\over{\sqrt \alpha}})X\mid X\in {\mathfrak g}\}$,
$\{(1-{\varepsilon\over{\sqrt \alpha}})X\mid X\in {\mathfrak g}\}$
commute with one another and both are isomorphic to ${\mathfrak
g}$. In fact, ${\mathfrak G}/{\mathfrak G}_{\alpha} \cong {\mathfrak
g}\oplus {\mathfrak g}$ with the natural Lie bracket:
$$[(X_{1},X_{2}),(Y_{1},Y_{2})] = ([X_{1},Y_{1}],[X_{2},Y_{2}])$$ with
the isomorphism given by $X + Y\varepsilon \mapsto (X + Y, X - Y)$.
Hence, the
double of a factorizable Lie bialgebra is isomorphic to ${\mathfrak
g}\oplus {\mathfrak g}$ and $D=G\times G$.

As deduced from (\ref{YangBaxter}) $r_{\pm}={\frac{1}{2}}(r \pm
\sqrt{\alpha})$ are Lie algebra morphisms from ${\mathfrak g}_{r}$ to
${\mathfrak g}$, i.e.
$$r_{\pm}[X,Y]_{r} = [r_{\pm}X,r_{\pm}Y]$$ and we have the following
embeddings of ${\mathfrak g}$ and ${\mathfrak g}_{r}$ in ${\mathfrak
g}\oplus {\mathfrak g}$:
$${\mathfrak g}_d = \{(X,X)\mid X\in {\mathfrak g}\},\quad {\mathfrak
g}_r = \{(r_{+}X,r_{-}X) \mid X \in {\mathfrak g}\}$$

We shall use the same notation ${\mathfrak g}_r$ for $({\mathfrak
g},[\cdot,\cdot]_r)$ and for its embedding in ${\mathfrak g}\oplus
{\mathfrak g}$. This should not lead to any confusion.

Notice that the map $X \mapsto (r_{+}X,r_{-}X)$ is non-degenerate as
long as $\alpha \neq 0$. We can recover $X$ by the formula
$X={\alpha}^{-{\frac{1}{2}}}(r_{+}X-r_{-}X)$. We shall often use the
notation $X_{\pm}=r_{\pm}X$.

Notice that ${\mathfrak g}_{\pm}:= r_{\pm}{\mathfrak g}$ are Lie
subalgebras of ${\mathfrak g}$ and denote by $G_{\pm}$ the subgroups
of $G\times G$ integrating ${\mathfrak g}_{\pm}$. We have the
following embeddings of $G$ and $G^*$:
$$G_d = \{(g,g)\in D\vert g\in G\},\quad G_r = \{(g_+,g_-)\in D\vert
g_+\in G_+,g_-\in G_-\}$$


The dressing transformations are given by the solutions of the
factorization problem:
\begin{equation}
(h_+,h_-)(g,g)=({\tilde g},{\tilde g})({\tilde h}_+,{\tilde h}_-)
\end{equation}

We can write now explicitly the Poisson-Lie structure on $G_{r}$ dual
to (\ref{PoissonLie}). After contraction with the right-invariant forms
on $G_{r}$, $\theta^{r}_R(X)=\tr[(\dd g_{+}g_{+}^{-1}- \dd g_{-}g_{-}^{-1})X]$
for $X \in {\mathfrak g}$, it takes the form
\begin{eqnarray} \label{dualPoissonLie}
P_{(g_{+},g_{-})}^{r}(X,Y)=\tr\left(X({\Ad}_{g_{+}}-{\Ad}_{g_{-}})
(r_{-}{\Ad}_{g_{+}}^{-1}-r_{+}{\Ad}_{g_{-}}^{-1})Y\right)
\end{eqnarray}
which verifies, in particular, that its linearization at the unit of
$G_r$ gives the Lie bracket of $\mathfrak g$.

Using the explicit description of the double group we can write the
symplectic structure (\ref{HeisenbergDouble}) at a point
$(h_{+}g,h_{-}g)=({\tilde g}{\tilde h}_{+}, {\tilde g}{\tilde
h}_{-})\in D_{0}$ as:
\begin{eqnarray}\label{HeiDoubleFact}
&&\Omega ((h_{+}g,h_{-}g)) ={\rm tr}\left(\dd\tilde{g}\tilde{g}^{-1}\wedge 
(\dd h_{+}h_{+}^{-1}-
\dd h_{-}h_{-}^{-1})+g^{-1}\dd g\wedge (\tilde{h}_{+}^{-1}\dd\tilde{h}_{+}-
\tilde{h}_{-}^{-1}\dd\tilde{h}_{-})\right)\notag
\end{eqnarray}

\newpage

{\it Example:}

Let ${\mathfrak g}$ be a simple Lie algebra over ${\mathbb C}$ and
$\phi$ its set of roots. The decomposition in root spaces reads
\begin{eqnarray} \label{decomposition}
{\mathfrak g} = {\mathfrak t} \oplus \bigg(\bigoplus_{\alpha \in \phi}
{\mathfrak g}_{\alpha}\bigg)
\end{eqnarray}
where ${\mathfrak t}$ is a Cartan subalgebra of ${\mathfrak g}$ and
$${\mathfrak g}_{\alpha}=\{{\mathbb C}X_{\alpha}\mid
[T,X_{\alpha}]=\alpha(T)X_{\alpha},\forall T\in {\mathfrak t}\}$$
Given a splitting into positive and negative roots, $\phi = \phi_+
\cup \phi_-$, any element $X \in {\mathfrak g}$ can be written as $X =
X^{(+)} + X^{(-)} + T$ where $X^{(\pm)} \in {\rm Span}(X_\alpha),\
\alpha \in \phi_{\pm}$.

The {\bf standard $r$-matrix} is defined by $r=r_+ + r_-$ with
\begin{eqnarray} \label{standardr}
&&r_+ X = X^{(+)}+\frac{1}{2}T \\
&&r_- X = -X^{(-)}-\frac{1}{2}T
\end{eqnarray}
which is a factorizable $r$-matrix with $\alpha=1$.

Take as a particular case ${\mathfrak g}={\mathfrak sl}(n,{\mathbb
C})$ with the standard $r$-matrix. Then, ${\mathfrak sl}(n,{\mathbb
C})_r \subset {\mathfrak sl}(n,{\mathbb C})\oplus {\mathfrak
sl}(n,{\mathbb C})$ consists of pairs $(X_+,X_-)$ where $X_+$ (resp.$
X_-$) is upper (resp. lower) triangular and ${\rm diag}(X_+)=-{\rm
diag}(X_-)$.

At the group level, $SL(n,{\mathbb C})_r \subset SL(n,{\mathbb
C})\times SL(n,{\mathbb C})$ is the set of pairs $(g_+,g_-)$ such that
$g_+$ (resp. $g_-$) is upper (resp. lower) triangular and ${\rm
diag}(g_+)={\rm diag}(g_-)^{-1}$.

\subsection{Triangular Lie bialgebras}


In this subsection we describe the double of a triangular Lie
bialgebra and the double of the associated Poisson-Lie groups. We
shall use these results for writing explicitly the Poisson structure
dual to (\ref{PoissonLie}) and, in the subsequent sections, for solving
the corresponding Poisson-Lie sigma models.

If $\alpha = 0$, $r_{\pm}$ degenerate to ${\frac{1}{2}}r$, the map $X
\mapsto ({\frac{1}{2}}rX,{\frac{1}{2}}rX)$ is not invertible and ${\mathfrak
G}/{\mathfrak G}_{0}$ is no longer isomorphic to ${\mathfrak g}\oplus
{\mathfrak g}$. Indeed, ${\mathfrak G}/{\mathfrak G}_{0}=\{X +
Y\varepsilon \mid {\varepsilon}^2 = 0\}$ is not semisimple, for the
elements $X\varepsilon$ form an abelian ideal as seen from the Lie
bracket
\begin{eqnarray} \label{bracket}
[X+Y\varepsilon,X'+Y'\varepsilon]=[X,X']+([X,Y']+[Y,X'])\varepsilon
\end{eqnarray}

This is the Lie algebra of the tangent bundle of $G$, $TG\cong G\times
{\mathfrak g}$, with the natural group structure given by the semidirect
product:
\begin{eqnarray} \label{semidirect}
(g,X)(g',X')=(gg',\Ad_{g}X'+X)
\end{eqnarray}

Hence, the double of a triangular Lie bialgebra is isomorphic to the
tangent bundle of $G$ with the product given by (\ref{semidirect}).

We can represent the elements of the double as
\begin{eqnarray}
D= \left\{ \begin{pmatrix}e&0 \\ X&e\end{pmatrix} g \mid g\in G,\ X\in {\mathfrak g}\right\}
\end{eqnarray}
where the product is now the formal product of matrices, resulting the
semidirect product mentioned above. Its Lie algebra with this notation is
\begin{eqnarray}
{\mathfrak d}=\left\{ \begin{pmatrix}X&0 \cr Y&X \end{pmatrix} \mid  X,Y \in {\mathfrak g}\right\}
\end{eqnarray}
and the Lie bracket (\ref{bracket}) is given by the formal commutator
of matrices. The embeddings of ${\mathfrak g}$ and ${\mathfrak g}_r$
in ${\mathfrak d}$ are given by:
\begin{eqnarray}
{\mathfrak g}_d = \left\{ \begin{pmatrix}X&0 \cr 0&X\end{pmatrix} \mid  X \in {\mathfrak g}\right\},\quad {\mathfrak g}_r = \left\{ \begin{pmatrix}rX&0 \cr X&rX \end{pmatrix} \mid  X \in {\mathfrak g}\right\}
\end{eqnarray}


Both subalgebras exponentiate to subgroups $G_d,G_{r} \subset D$. 
Clearly,
\begin{eqnarray}
G_d=\left\{ \begin{pmatrix}g&0 \cr 0&g \end{pmatrix}\mid g\in G \right\}
\end{eqnarray}
whereas for $G_{r}$ the description is less explicit. It is the subgroup of $D$ generated by elements of the form:
\begin{eqnarray} \label{Gr}
{\begin{pmatrix}e&0 \cr Y&e \end{pmatrix}}e^{rX} \quad {\rm with}\quad
Y=\int^{1}_{0}\Ad_{e^{srX}}X \dd s,\ X\in {\mathfrak g}.
\end{eqnarray}
We will denote a general
element of $G_r$ by
\begin{eqnarray}
{\bar Y}={\begin{pmatrix}e&0 \cr Y&e \end{pmatrix}}h_Y
\end{eqnarray}
where, in the general case, $Y$ belongs to a dense subset of ${\mathfrak g}$ and determines
$h_{Y}$ up to multiplication by an element of $G_0$. This means that
$${\rm if} \quad  {\begin{pmatrix}e&0 \cr Y&e \end{pmatrix}}h_Y \in G_{r},\ {\rm then}$$
$${\begin{pmatrix}e&0 \cr Y&e\end{pmatrix}}h'_Y \in G_{r} \Leftrightarrow h'_{Y} = h_{Y}g_{0}, \ g_{0} \in G_{0}.$$
As a consequence, the notation ${\bar Y}$ has a small ambiguity, but
we shall use of it for brevity wherever it does not lead to
confusion.

The realization of $G^*$ described above allows us to write the Poisson structure dual to (\ref{PoissonLie}).

The right-invariant forms in $G_{r}$ are
$$\theta^r_R(Y)=\tr\left(\left(\dd X+[X,\dd
h_{X}h_{X}^{-1}]\right)Y\right)=\tr\left(\dd
\left(\Ad_{h_X}^{-1}X\right)\Ad_{h_X}^{-1}Y\right)$$
for $Y\in{\mathfrak g}$ and $\begin{pmatrix}e&0\cr X&e\end{pmatrix}h_X\in G_r$.
Whereas the left invariant forms read:
$$\theta^r_L(Y)=\tr\left(Y\Ad^{-1}_{h_X}\dd X\right).$$
It can be checked after a straightforward (although lengthy)
calculation that the dual Poisson-Lie structure contracted with the
right-invariant forms is
\begin{eqnarray} \label{dualPoissonLieTriangular}
P_{\bar X}^{r}(Y,Z)=\tr\left(Y[X,Z]-[X,Y]\Ad_{h_X}r\Ad_{h_X}^{-1}[X,Z]\right)\end{eqnarray}

The symplectic structure on the Heisenberg double
(\ref{HeisenbergDouble}) at a point ${\bar X}g = {\tilde
g}{\tilde{\bar X}} \in D_0$ can be written now:
\begin{equation}
\Omega ({\bar X}g) = \tr \left(\dd {\tilde g} {\tilde g}^{-1}\wedge(\dd X+[X,\dd
h_X h_X^{-1}]) + g ^{-1}\dd g\wedge \Ad_{h_{\tilde X}}^{-1}\dd {\tilde X}
\right)
\end{equation}

\vskip 0.4 cm

{\it Example:}

Take ${\mathfrak g}$ a complex simple Lie algebra. If $\tau_{\mathfrak
t}:{\mathfrak g}\rightarrow {\mathfrak t}$ is the projector onto the
Cartan subalgebra ${\mathfrak t}$ with respect to the decomposition
(\ref{decomposition}) and ${\cal O}:{\mathfrak t}\rightarrow
{\mathfrak t}$ is an antisymmetric endomorphism of ${\mathfrak t}$
with respect to ${\rm tr}(\ )$, then $r={\cal
O}\tau_{\mathfrak t}$ is an $r$-matrix with $\alpha = 0$.

It is worth studying in detail the structure of $G_{r}$ for this
$r$. As we know, $G_{r}$ is generated by elements of the form
\begin{eqnarray} \label{dualgroup}
\begin{pmatrix}e&0\cr Y&e \end{pmatrix}e^{rX}, 
\ Y=\int^{1}_{0}\Ad_{e^{srX}}X\dd s,\ X\in {\mathfrak g}
\end{eqnarray}

The elements $\{rX\mid X\in{\mathfrak g}\}$ span a subalgebra of
${\mathfrak t}$ (therefore abelian) and its exponentiation will be a
subgroup of the Cartan subgroup of $G$, so we concentrate on $Y$.

Take $$X=T + \sum_{\alpha \in \phi} a_{\alpha}X_{\alpha}, \ T \in
{\mathfrak t}$$

Using that
$$\Ad_{e^{srX}}=e^{ad(srX)}$$ we straightforwardly obtain
\begin{eqnarray}
\Ad_{e^{srX}}X = T + \sum_{\alpha \in \phi} e^{s\alpha
(rT)}a_{\alpha}X_{\alpha}
\end{eqnarray}
and, therefore,
\begin{eqnarray}
Y = T + \sum_{\alpha \in \phi}{1\over{\alpha(rT)}}\left(e^{\alpha
(rT)}-1\right)a_{\alpha}X_{\alpha}
\end{eqnarray}
where, if some $\alpha(rT)=0$, the limit $\alpha(rT)\rightarrow 0$
must be understood in the last expression.

The product of $n$ elements of the form (\ref{dualgroup})
$$\begin{pmatrix}e&0\cr Y&e \end{pmatrix}T_{Y} = \begin{pmatrix}e&0\cr
Y_{1}&e \end{pmatrix}e^{rX_{1}}\dots \begin{pmatrix}e&0\cr Y_{n}&e
\end{pmatrix}e^{rX_{n}}$$
yields
\begin{eqnarray} \label{generaldualgroup}
T_{Y} &=& e^{rT_{1} + \dots + rT_{n}}\cr Y &=& T_{1}+ \dots +T_{n} +
\sum_{\alpha \in \phi}\bigg[{1\over{\alpha(rT_{1})}}\left(e^{\alpha
(rT_{1}+ \dots +rT_{n})}-e^{\alpha (rT_{2}+ \dots
+rT_{n})}\right)a_{\alpha,1} +\cr &+&
{1\over{\alpha(rT_{2})}}\left(e^{\alpha (rT_{2}+ \dots
+rT_{n})}-e^{\alpha (rT_{3} + \dots +rT_{n})}\right)a_{\alpha,2} + \cr
&+& \dots + {1\over{\alpha(rT_{n})}}\left(e^{\alpha
(rT_{n})}-1\right)a_{\alpha,n}\bigg]X_{\alpha}
\end{eqnarray}
where
$$X_i=T_i + \sum_{\alpha \in \phi} a_{\alpha ,i}X_{\alpha}, \ T_i \in
{\mathfrak t}$$

{}From (\ref{generaldualgroup}) it is clear that, in this case,
$G_{0}=e$ and that $Y$ fills ${\mathfrak g}$. As a consequence, the
dressing actions are globally defined, $D=GG_{r}=G_{r}G$ and the
factorizations are unique.

\vskip 0.8 cm

{\it Example:}

$r=0$ is an $r$-matrix with $\alpha = 0$. It endows $G$ with trivial
Poisson bracket $\{\cdot,\cdot\}_G = 0$ and ${\mathfrak g}^*$ with trivial Lie
bracket $[\cdot,\cdot]_{{\mathfrak g}^*}=0$. The dual Poisson Lie group $G^*$ is
${\mathfrak g}^*$ viewed as an abelian group and equipped with the
so-called {\bf Kostant-Kirillov Poisson bracket}, namely:
\begin{eqnarray}
\{X,Y\}_{{\mathfrak g}^*}(\xi)=\left<\xi,[X,Y]\right>,\ X,Y \in {\mathfrak g},\ \xi \in {\mathfrak g}^*
\end{eqnarray}
for linear functions on ${\mathfrak g}^*$ (elements of ${\mathfrak g}$) and extended by the Leibniz rule to $C^\infty({\mathfrak g}^*)$.

\section{The Poisson-Sigma model} \label{IntroPS}

The Poisson-Sigma model (\cite{SchStr94}) is a two-dimensional
topological sigma model with a Poisson manifold $(M,\Pi)$ as
target. The fields of the model are $X:\Sigma \rightarrow M$ and a
1-form $\psi$ on $\Sigma$ with values in the pull-back by $X$ of the
cotangent bundle of $M$.  The action functional has the form
\begin{eqnarray} \label{PS}
S_{P\sigma}(X,\psi)=\int_\Sigma \langle \dd X,\wedge\psi\rangle - 
{\frac{1}{2}}\langle\Pi\circ X,\psi\wedge\psi\rangle
\end{eqnarray}
where $\langle\cdot ,\cdot \rangle$ denotes the pairing between
vectors and covectors of $M$.

If $X^{i}$ are local coordinates on $M$, $\sigma ^{\mu}$ local
coordinates on $\Sigma$, $\Pi^{ij}(X)$ the components of the Poisson
structure in these coordinates and
$\psi_{i}=\psi_{i\mu}d{\sigma}^{\mu}$, $i=1,...,n; \mu=1,2$ the action
reads
\begin{eqnarray} \label{PScoor}
S_{P\sigma}(X,\psi)=\int_\Sigma \dd X^{i}\wedge \psi_{i}-
{\frac{1}{2}}\Pi^{ij}(X)\psi_{i}\wedge \psi_{j}
\end{eqnarray}

The equations of motion in the bulk are:
\begin{subequations}\label{eom}
\begin{align}
&\dd X^{i}+\Pi^{ij}(X)\psi_{j}=0 \label{eoma} \\ 
&\dd \psi_{i}+{\frac{1}{2}}\partial_{i}\Pi^{jk}(X)\psi_{j}\wedge \psi_{k}=0 \label{eomb}
\end{align}
\end{subequations}

In particular, $X$ lies within one of the symplectic leaves of the
foliation of $M$.

Take $\epsilon=\epsilon_{i}\dd X^i$ a section of $X^*T^*M$. The
infinitesimal transformation
\begin{eqnarray} \label{symmetry}
&&\delta_{\epsilon}X^{i}=\epsilon_{j}\Pi^{ji}(X)\cr
&&\delta_{\epsilon}\psi_{i}=d\epsilon_{i}+\partial_{i}\Pi^{jk}(X)\psi_{j}
\epsilon_{k}
\end{eqnarray}
leaves the action (\ref{PScoor}) invariant up to a boundary term
\begin{eqnarray} \label{symmS}
\delta_{\epsilon}S_{P\sigma}=-\int_\Sigma \dd(\dd X^i \epsilon_i).
\end{eqnarray}

Notice that the gauge transformations (\ref{symmetry}) form an open
algebra, as the commutator of two of them closes only on-shell:
\begin{subequations} \label{commg}
\begin{align}
 &[\delta_\epsilon,\delta_{\epsilon'}]X^i=\delta_{[\epsilon,\epsilon']^*} X^i \label{commga} \\
&[\delta_\epsilon,\delta_{\epsilon'}]\psi_i=\delta_{[\epsilon,\epsilon']^*} \psi_i
+\epsilon_k\epsilon_{l}'\partial_i\partial_j \Pi^{kl}(\dd X^{j}+\Pi^{js}(X)\psi_{s})\label{commgb}
\end{align}
\end{subequations}
where $[\epsilon,\epsilon']^*_k=
\partial_k\Pi^{ij}(X)\epsilon_i\wedge\epsilon'_j$. 

The general study of the model defined on a surface with boundary was
carried out in \cite{CalFal04} and \cite{CalFal051}. Assume that $X$
at the boundary of $\Sigma$ is restricted to a submanifold (brane) $N
\hookrightarrow M$. Define ${\cal I}_N :=\{f \in C^\infty(M) \vert
f(p) = 0,\ p \in N\}$ and ${\cal F}_N:=\{f\in C^\infty(M) \vert
\{f,{\cal I}_N\}\subset{\cal I}_N\}$. The brane $N$ is {\bf
classically admissible}\footnote{The quantization of the model on the
disk with a general brane was discussed in \cite{CalFal051}. Therein,
a regularity condition stronger than (\ref{weakregularity}) 
was
imposed to the brane.} 
if it
satisfies the regularity condition
\begin{eqnarray} \label{weakregularity}
\dim\{(\dd f)_p | f\in{\cal F}_N\cap{\cal I}_N\}={\rm const.},\ \forall
p\in N
\end{eqnarray}

If $N$ is classically admissible the Poisson-Sigma model with the
boundary condition $X\vert_{\partial\Sigma} \in N$ is consistent. The
appropriate boundary condition for $\psi$ is that its contraction with
vectors tangent to $\partial \Sigma$, $\psi_{t}$, takes values in
$\{\dd f | f\in{\cal F}_N\cap{\cal I}_N\}$, and the gauge transformations
are restricted at the boundary to the same space.

\section{Poisson-Lie sigma models}

In this section we recall the results of (\cite{CalFalGar}) for
factorizable Poisson-Lie groups and treat in detail the triangular
case.

 When the target manifold is a Lie group, the action of the
 Poisson-Sigma model can be recasted in terms of a set of fields
 adapted to the group structure. $T^{*}G$ can be identified, by right
 translations, with $G\times{\mathfrak g}^*$ and, using ${\rm tr}(\ )$,
 with $G \times {\mathfrak g}$. Then, in (\ref{PS}) we can take
 $A\in\Lambda^1(\Sigma)\otimes{\mathfrak g}$ (instead of $\psi$) and
 $g:\Sigma\rightarrow G$ as fields and use the Poisson structure
 contracted with the right-invariant forms in $G$
 (\ref{PoissonContracted}). Denoting by $P_g^\sharp:{\mathfrak
 g}\rightarrow {\mathfrak g}$ the endomorphism induced by $P_g$ using
 ${\rm tr}(\ )$ we can write the action of the Poisson-Sigma model as
 
\begin{equation} \label{PSgroup}
S(g,A)=\int_{\Sigma}{\tr(\dd gg^{-1}\wedge A)-{\frac{1}{2}}\tr(A\wedge
P_g^\sharp A)}
\end{equation}

In particular, for the Poisson-Lie structure (\ref{PoissonLie}) we have
\begin{eqnarray} \label{PLS}
S_{PL}(g,A)=\int_{\Sigma}{\tr(\dd gg^{-1}\wedge A)-{1\over4}\tr(A\wedge
(r-\Ad_{g}r\Ad_{g}^{-1})A)}
\end{eqnarray}
which is the action of what we shall call {\bf Poisson-Lie sigma
model} with target $G$.

The equations of motion are
\begin{subequations} \label{eomG}
\begin{align}
&\dd gg^{-1}+{\frac{1}{2}}(r-\Ad_{g}r\Ad_{g}^{-1})A=0\label{eomGa} \\
&\dd {\tilde A}+[{\tilde A},{\tilde A}]_{r}=0,\ {\tilde A}:=\Ad_{g}^{-1}A
\label{eomGb}
\end{align}
\end{subequations}
from which a zero curvature equation can be also derived for $A$,
\begin{eqnarray}\label{Aflat}
&&\dd A+[{A},{A}]_{r}=0,
\end{eqnarray}

The infinitesimal gauge symmetry of the action, for
$\beta:\Sigma\rightarrow {\mathfrak g}$ is
\begin{subequations} \label{gaugesymmetryG}
\begin{align}
&\delta_{\beta}g g^{-1}={\frac{1}{2}}(\Ad_{g}r\Ad_{g}^{-1}-r)\beta 
\label{gaugesymmetryGa} \\
&\delta_{\beta}A=\dd \beta +
[A,\beta]_{r}-{\frac{1}{2}}[\dd gg^{-1}+{\frac{1}{2}}(r-\Ad_{g}r\Ad_{g}^{-1})A,\beta]
\label{gaugesymmetryGb}
\end{align}
\end{subequations}
which corresponds to the right dressing vector fields of \cite{LuWei}
translated to the origin by right multiplication in $G$. Its
integration (local as in general the vector field is not complete)
gives rise to the dressing transformation of $g$. On-shell,
$[\delta_{\beta_{1}},\delta_{\beta_{2}}]=\delta_{[\beta_{1},\beta_{2}]_{r}}$
and we can talk properly about a gauge group. In fact, the gauge group
of the Poisson-Sigma model on $G$ is its dual $G^*$.

Up to here we have not needed to distinguish between $\alpha=0$ and
$\alpha\neq 0$. In order to study further (\ref{PLS}) and its dual
model we need to make use of the embeddings of $G$ and $G^{*}$ in the
double $D$. As we have learnt, $D$ is very different in the
factorizable and triangular cases and we must analyse them
separately.

\subsection{Factorizable Poisson-Lie sigma models}

It was shown in (\cite{CalFalGar}) that locally the solutions in the bulk are:

\begin{equation}
A=h_{+}\dd h_{+}^{-1}-h_{-}\dd h_{-}^{-1}
\end{equation}
and $g(\sigma)$ is given by the solution of
\begin{eqnarray} \label{solFact}
(h_+(\sigma)\hat g,h_-(\sigma)\hat g)=
(g(\sigma){\tilde h}_+(\sigma),g(\sigma)\tilde h_-(\sigma))
\end{eqnarray}

We go on to study the reduced phase space of the model when
$\Sigma={\mathbb R}\times [0,\pi]$ and $g$ is free at the
boundary. Equivalently, in the language of Section 4, we take a
brane which is the whole target manifold $N=G$. The $A$ field must
then vanish on vectors tangent to $\partial\Sigma$, so that $h_\pm,
{\tilde h}_\pm$ are constant along the connected components of the
boundary. Writing $\sigma=(t,x)$, we have $h_\pm(t,0)=h_{0\pm}$,
${\tilde h}_{\pm}(t,0)={\tilde h}_{0\pm}$, $h_\pm(t,\pi)= h_{\pi\pm}$,
${\tilde h}_\pm(t,\pi)={\tilde h}_{\pi\pm}$.

Denote $I=[0,\pi]$. The canonical symplectic form on the space of
continuous maps $(g,A_x):TI\rightarrow G\times {\mathfrak g}$ with
continuously differentiable base map is:
$$\omega =\int_0^\pi \tr(\delta g g^{-1}\wedge \delta g
g^{-1}A_x-\delta g g^{-1}\wedge \delta A_x)\dd x$$

When restricted to the solutions of the equations of motion the
symplectic form $\omega$ becomes degenerate, its kernel given by the
gauge transformations (\ref{gaugesymmetryG}) which vanish at
$x=0,\pi$. By definition, the reduced phase space\footnote{${\cal
P}(M,N)$ stands for the reduced phase space of the Poisson-Sigma model
with target $M$ and brane $N$.} ${\cal P}(G,G)$ is the (possibly
singular) quotient of the space of solutions by the kernel of
$\omega$.

If we parametrize the solutions in terms of 
$h_\pm(\sigma), {\hat g}$
we obtain
\begin{eqnarray}
\omega={\frac{1}{2}}\int_0^\pi \partial_x\Omega((h_+(\sigma)\hat
g,h_-(\sigma)\hat g))\dd x
\end{eqnarray}
That is, $\omega$ depends only on the values of the fields at the
boundary (i.e. the degrees of freedom of the theory are all at the
boundary, as expected from the topological nature of the model) and is
expressed in terms of the symplectic structure on the Heisenberg
double (\ref{HeiDoubleFact}). Namely,
$$\omega={\frac{1}{2}}[\Omega((h_{\pi+}\hat g,h_{\pi-}\hat g))-
\Omega((h_{0+}\hat g,h_{0-}\hat g))]$$

Or if we take $\sigma_0=(t_0,0)$, i.e. $h_{0\pm}={\tilde h}_{0\pm}=e$
$$\omega={\frac{1}{2}}\Omega((h_{\pi+}\hat g,h_{\pi-}\hat g))$$

The reduced phase space ${\cal P}(G,G)$ is then the set of pairs
$([(h_+,h_-)],\hat g)$ with $[(h_+,h_-)]$ a homotopy class of maps
from $[0,\pi]$ to $G_r$ which are the identity at $0$ and have fixed
value at $\pi$ and such that $(h_+(x)\hat g,h_-(x)\hat g)\in D_0,\
x\in[0,\pi]$. The symplectic form on ${\cal P}(G,G)$ can be viewed as
the pull-back of $\Omega$ by the map $([h_{+},h_{-}],{\hat g})\mapsto
(h_{\pi +}{\hat g},h_{\pi -}{\hat g})$.

\subsubsection{The dual factorizable model}

Using the Poisson structure given in (\ref{dualPoissonLie}) the action
of the dual model reads
\begin{eqnarray}\label{dualPLS}
S_{PL}^{*}(g_{+},g_{-},A)&=&\int_\Sigma \tr[(\dd\gp\gpinv
-\dd\gm\gminv)\wedge A+ \cr
&&+{\frac{1}{2}}A\wedge(\Ad_{\gp}-\Ad_{\gm})(r_+\Ad_{\gm}^{-1}
-r_-\Ad_{\gp}^{-1})A]\end{eqnarray}

As shown in (\cite{FalGaw02},\cite{AleSchStr}) the Poisson-Lie sigma
model with target $G_{r}$ and fields $(g_{+},g_{-})$ and $A$ is
locally equivalent to the $G/G$ WZW model with fields
$g=g_{-}g_{+}^{-1}$ and $A$. This relation can be established for any
factorizable $r$-matrix.

The equations of motion of the model can be written
\begin{eqnarray}\label{eqofmotdual}
&&g_{\pm}^{-1}\dd g_\pm+r_{\pm}({\Ad}_{g_+}^{-1}-{\Ad}_{g_-}^{-1})A=0\cr\cr
&&\dd{A}+[{A}, {A}]=0
\end{eqnarray}

The gauge transformations, for $\beta:\Sigma\rightarrow{\mathfrak g}$, read:
\begin{subequations} \label{symmdual}
\begin{align}
&\hskip-.7cm g_{\pm}^{-1}\delta_\beta
g_\pm=r_{\pm}({\Ad}_{g_-}^{-1}-{\Ad}_{g_+}^{-1})\beta \label{symmduala}
\\ \notag
\\ 
&\hskip -.7cm\delta_\beta
A=\dd\beta+[A,\beta]+ \notag \\
&+{\frac{1}{2}}
(r_+{\Ad}_{g_-}+r_-{\Ad}_{g_+})[g_{+}^{-1}\dd g_+-g_{-}^{-1}\dd g_-+
({\Ad}_{g_+}^{-1}-{\Ad}_{g_-}^{-1})A,\tilde\beta] \label{symmdualb}
\end{align}
\end{subequations}
where $\tilde\beta:=(r_- {\Ad}_{g_+}^{-1}-r_+ {\Ad}_{g_-}^{-1})\beta$.  
The gauge transformations
close on shell. Namely, 
$[\delta_\beta,\delta_\gamma]=\delta_{[\beta,\gamma]}$
which corresponds now to the gauge group $G$.

The solutions of the equations of motion can be obtained along
the same lines as before. Locally,
\begin{eqnarray}
A=h\dd h^{-1}
\end{eqnarray}
And $(g_+(\sigma),g_-(\sigma))$ is obtained as the solution of:
$$
(g_+(\sigma){\tilde h}(\sigma)
,g_-(\sigma){\tilde h}(\sigma))=
({h}(\sigma)\hat g_+,
{h}(\sigma)\hat g_-),
$$ which means that $(g_+,g_-)$ is the dressing-transformed of $(\hat
g_+,\hat g_-)$ by $h$. 
At this point it is evident the symmetry
between both dual models under the exchange of the roles of $G$ and
$G_r$.

In the open geometry with free boundary conditions
$h$ is constant along connected components of the boundary and 
one may take 
$h(t,0)=\tilde
h(t,0)=e$, $h(t,\pi)=h_\pi$, 
$\tilde h(t,\pi)=\tilde h_\pi$.
The symplectic form 
can then be written
\begin{eqnarray}
\omega^{*}={\frac{1}{2}}\Omega(({h_\pi}\hat g_+,{h_\pi}\hat g_-)).
\end{eqnarray}

The duality between ${\cal P}(G,G)$ and ${\cal P}(G_r,G_r)$ was
pointed out in \cite{CalFalGar}. The symplectic forms of the two
models coincide upon the exchange of $h_\pi$ with $\hat g^{-1}$ and
$(\hat g_+,\hat g_-)$ with $(h_{\pi+}^{-1},h_{\pi-}^{-1})$. Hence, one
can talk about a bulk-boundary duality between the Poisson-Lie sigma
models for $G$ and $G^*$ since the exchange of degrees of freedom maps
variables associated to the bulk of one model to variables associated
to the boundary of the other one.

In the next subsection we make use of the explicit realization of the
double in the triangular case given in Section 3.2 for solving the
corresponding Poisson-Lie sigma models. We shall see that the
bulk-boundary duality found in the factorizable case still holds.

\subsection{Triangular Poisson-Lie sigma models}

We now go back to equations (\ref{eqofmotdual}) and assume
$r$ is triangular ($\alpha = 0$). We start noting that whereas $A$ is
a pure gauge of the group $G_r$, ${\frac{1}{2}}rA$ is a pure gauge of
the group $G$, i.e.
$$\dd ({\frac{1}{2}}rA)+[{\frac{1}{2}}rA,{\frac{1}{2}}rA]=0$$

Now take
$${\bar X}(\sigma) = \begin{pmatrix}e&0\cr X(\sigma)&e\end{pmatrix}h_{X}(\sigma) \in G_{r}$$

Then, locally,
\begin{eqnarray}
{\bar X}^{-1}\dd{\bar X} = \begin{pmatrix}h_{X}^{-1}\dd h_{X}&0 \cr
h_{X}^{-1}\dd Xh_{X}&h_{X}^{-1}\dd
h_{X}\end{pmatrix}=\begin{pmatrix}rA&0 \cr A&rA\end{pmatrix}
\end{eqnarray}

${\tilde A}$ is also a pure gauge of the group $G_r$, so ${\tilde
A}=h_{{\tilde X}}^{-1}d{\tilde X}h_{{\tilde X}}$ and the equation of
motion for $g$ reads,
\begin{eqnarray}
\dd gg^{-1}+{\frac{1}{2}}(h_{X}^{-1}\dd h_{X}-\Ad_{g}h_{{\tilde
X}}^{-1}\dd h_{{\tilde X}})=0
\end{eqnarray}
which implies
\begin{eqnarray}
g=h_{X}^{-1}{\hat g}h_{\tilde X}, \ {\hat g}\in G
\end{eqnarray}

${\tilde A}=\Ad_{g}^{-1}A \Rightarrow {\tilde X}=\Ad_{\hat g}^{-1}X$ and
we can write, locally, the solution as an equation in the double:
\begin{eqnarray} \label{solutiondirect}
\begin{pmatrix}e&0\cr X&e\end{pmatrix}h_{X}\begin{pmatrix}e&0\cr 0&e\end{pmatrix}g = \begin{pmatrix}e&0 \cr 0&e\end{pmatrix}{\hat g}\begin{pmatrix}e&0\cr {\tilde X}&e\end{pmatrix}h_{\tilde X}
\end{eqnarray}

Therefore, in the language of dressing actions, $g={}^{{\bar
X}^{-1}}{\hat g}$.

The analysis of the reduced phase space when $\Sigma={\mathbb
R}\times[0,\pi]$ and $g\vert_{\partial\Sigma}$ is free works as in the
factorizable case. The field $A$ must vanish on vectors tangent to
${\partial\Sigma}$ and hence $\bar X$ is constant along each connected
component of the boundary.

By using the explicit solution (\ref{solutiondirect}) we
can identify ${\cal P}(G,G)$. Notice that we can always choose ${\bar
X}(t,0)={\bar{\tilde X}}(t,0)=e$. With this choice and defining
$X_{\pi}:=X(t,\pi),{\tilde X}_{\pi}:={\tilde
X}(t,\pi),g_{\pi}:=g(t,\pi)$, a straightforward calculation yields
\begin{eqnarray}
\omega &=& {\frac{1}{2}}\tr \left(\delta X_{\pi}+[X_{\pi},\delta
h_{X_\pi}h_{X_\pi}^{-1}])\wedge \delta {\hat g} {\hat g}^{-1} +
\Ad_{h_{{\tilde X}_\pi}}^{-1}\delta {{\tilde X}_\pi} \wedge
g^{-1}_{\pi}\delta g_{\pi}\right)= \\
&=&{\frac{1}{2}}\Omega({\bar X}_\pi{\hat g})
\end{eqnarray}
 The reduced phase space ${\cal P}(G,G)$ turns out to be the set of pairs
$([{\bar X}],{\hat g})$ with $[{\bar X}]$ a homotopy class of maps
from $[0,\pi]$ to $G_r$ which are the identity at $x=0$ and have fixed
value at $x = \pi$.

\subsubsection{The dual triangular model and BF-theory}

Take (\ref{dualPoissonLieTriangular}) and write the action of the
Poisson-Sigma model with target $G_r$:
\begin{eqnarray}
S^{*}_{PL}({\bar X},A)&=&\int_{\Sigma}\tr \Big((\dd X\wedge A + \dd
X\wedge
\Ad_{g_{{}_X}}r\Ad_{g_{{}_X}}^{-1}[X,A]+{\frac{1}{2}}A\wedge[X,A]-\cr
&-&{\frac{1}{2}}[X,A]\wedge \Ad_{g_{{}_X}}r\Ad_{g_{{}_X}}^{-1}[X,A]\Big)
\end{eqnarray}
with fields $A \in \Lambda^{1}(\Sigma)\otimes {\mathfrak g}$, ${\bar
X}:\Sigma \to G_{r}$.

Note that the action is actually determined by
$X$ and $A$, since it is invariant under $g_{X} \mapsto g_{{}_X}g_{0},\
g_{0} \in G_{0}$.

Varying the action with respect to $A$ we get the equation of motion for $X$,
\begin{eqnarray} \label{eomX}
\dd X + [A,X] = 0
\end{eqnarray}

Taking variations with respect to ${\bar X}$ and after a rather cumbersome
 calculation we obtain the equation of motion for $A$,
\begin{eqnarray} \label{eomA}
\dd A + [A,A] = 0
\end{eqnarray}

The infinitesimal gauge symmetry for $\beta:\Sigma\rightarrow {\mathfrak g}$
\begin{subequations} \label{dualgaugesymmetry}
\begin{align}
&\delta_{\beta}X = [X,\beta] \label{dualgaugesymmetrya} \\
&\delta_{\beta}A=\dd \beta + [A,\beta]-r[\dd X+[A,X],\beta+\Ad_{g_{{}_X}}r\Ad_{g_{{}_X}}^{-1}[X,\beta]] \label{dualgaugesymmetryb}
\end{align}
\end{subequations}
which this time corresponds to the vector fields of the infinitesimal form of the right dressing action of $G$ on $G^{*}$. On-shell,$[\delta_{\beta_{1}},\delta_{\beta_{2}}]=\delta_{[\beta_{1},\beta_{2}]}$ 

The solutions of the equations of motion are, locally,
\begin{eqnarray} \label{solutiondual}
&&A=h^{-1}\dd h \cr
&&X=\Ad_{h}^{-1}{\hat X}
\end{eqnarray}
with $h:U \rightarrow G$, $U\subset \Sigma$ an open contractible subset and
${\hat X} \in {\mathfrak g}$.

This can also be written as an equation in $D$:
\begin{eqnarray} \label{solutiondualdouble}
\begin{pmatrix}e&0\cr {\hat X}&e\end{pmatrix}g_{\hat X}\begin{pmatrix}e&0\cr 0&e\end{pmatrix}{\tilde h} = \begin{pmatrix}e&0 \cr 0&e\end{pmatrix}h \begin{pmatrix}e&0\cr X&e\end{pmatrix}g_{{}_X}
\end{eqnarray}
which can be obtained from (\ref{solutiondirect}) taking $X\rightarrow
{\hat X}$, ${\hat g}\rightarrow h$, $g\rightarrow {\tilde h}$,
${\tilde Y}\rightarrow Y$. Now, ${\bar X}={}^{h^{-1}}{\hat{\bar X}}$.

Now consider $\Sigma={\mathbb R}\times [0,\pi]$ and ${\bar
X}\vert_{\partial\Sigma}$ free. $h$ must be constant along each
connected component of the boundary.

By choosing $h(t,0)={\tilde h}(t,0)=e$, defining
$h_{\pi}:=h(t,\pi),{\tilde h}_{\pi}:={\tilde
h}(t,\pi),X_{\pi}:=X(t,\pi)$ and plugging in (\ref{solutiondual}), we
get
\begin{eqnarray}
\omega^* &=& {\frac{1}{2}}\tr \left((\delta {\hat X}+[{\hat X},\delta {\hat
g_{{}_X}}{\hat g_{{}_X}}^{-1}]) \wedge \delta {\tilde h}_{\pi} {\tilde
h}_{\pi}^{-1} + \Ad_{g_{X_{\pi}}}^{-1}\delta {X_\pi} \wedge {\tilde h}
^{-1}_{\pi}\delta {\tilde h} _{\pi}\right)= \\
&=&{\frac{1}{2}}\Omega(h_\pi {\hat{\bar X}})
\end{eqnarray}

The reduced phase space ${\cal P}(G_r,G_r)$ is the set of pairs
$([h],{\hat {\bar X}})$ with $[h]$ a homotopy class of maps from
$[0,\pi]$ to $G$ which are the identity at $x=0$ and have fixed value
at $x = \pi$. Notice the duality between ${\cal P}(G,G)$ and ${\cal
P}(G_r,G_r)$ under the interchange ${\hat g} \leftrightarrow h_\pi,\
X_\pi \leftrightarrow {\hat X}$. This is the triangular version of the
bulk-boundary duality found in \cite{CalFalGar} and recalled in
Section 5.1 for the factorizable case.

\vskip 1 cm

Now, consider as target of the Poisson-Sigma model the dual of a
simple complex Lie algebra ${\mathfrak g}^*$ with the Kostant-Kirillov
Poisson bracket. As mentioned in Section 3.2 this is the dual
Poisson-Lie group of the simply connected Lie group $G$ whose Lie
algebra is ${\mathfrak g}$ endowed with the zero Poisson
structure. The action in this particular case is:

\begin{eqnarray} \label{BFtheory}
S_{{\rm BF}}=\int_{\Sigma}\tr \left(\dd X \wedge A
-{\frac{1}{2}}[X,A]\wedge A\right)
\end{eqnarray}
with $X:\Sigma \rightarrow {\mathfrak g}$ and $A \in
\Lambda^{1}(\Sigma)\otimes {\mathfrak g}$. This is the action of
BF-theory (\cite{Hor}) up to a boundary term.

The equations of motion are
\begin{eqnarray} \label{eomLinear}
&&\dd X+[A,X]=0 \cr
&&\dd A + [A,A] = 0
\end{eqnarray}

For $\epsilon\in \Lambda^0(M)\otimes{\mathfrak g}$ the gauge transformation
\begin{subequations}
\begin{align}
&\delta_\epsilon X=[X,\epsilon] \\
&\delta_\epsilon A= \dd \epsilon+ [A,\epsilon]
\end{align}
\end{subequations}
induces the change of the action (\ref{BFtheory}) by a boundary term
\begin{eqnarray}
\delta_\epsilon S_{{\rm BF}}=-\int_\Sigma \dd \tr (\dd X\epsilon).
\end{eqnarray}

Note that in this case the gauge transformations close even off-shell
$$[\delta_\epsilon,\delta_{\epsilon'}]=\delta_{[\epsilon,\epsilon']}$$
and induce the Lie algebra structure of $\Lambda^0(\Sigma)\otimes{\mathfrak g}$
in the space of parameters.

The equations of motion (\ref{eomX}), (\ref{eomA}) are the same as
(\ref{eomLinear}). We would like to understand this fact at the level
of the action. A direct computation shows that the following equality
holds:
\begin{eqnarray} \label{equivalence}
S^*_{PL}({\bar X},A)=S_{{\rm BF}}-{\frac{1}{2}}(\dd
X+[A,X])\Ad_{g_{{}_X}}r\Ad_{g_{{}_X}}^{-1}(\dd X+[A,X])
\end{eqnarray}

Hence, both actions differ by terms quadratic in the equations of
motion. This means that the Poisson-Lie sigma model with target $G_r$
is equivalent, for any triangular $r$-matrix, to the Poisson-Lie sigma
model over $G_{r=0}$, i.e. BF-theory. This is the triangular version
of the connection encountered in the factorizable case
(\cite{FalGaw02}), where every Poisson-Lie sigma model with target
$G_r$ for any factorizable $r$-matrix is (locally) equivalent to the
$G/G$ WZW model with target $G$.

\section{D-branes in Poisson-Lie sigma models}

In the previous section we have seen that the moduli spaces of
solutions of the Poisson-Lie sigma models over $G$ and $G_r\cong G^*$
coincide when $g$ is free at the boundary (i.e. when the brane is the
whole target group). We would like to find out whether such duality
holds for more general boundary conditions. That is, we address the
problem of finding pairs of branes $N\subset G$ and $N^*\subset G^*$
such that ${\cal P}(G,N)\cong {\cal P}(G^*,N^*)$.

Let us restrict $g\vert_{\partial\Sigma}$ to a closed submanifold
(brane) $N\subset M$. It is natural to ask the brane $N$ to respect
the Poisson-Lie structure of $G$ given by the $r$-matrix. To this end
we consider a simple subalgebra ${\mathfrak h}\subset {\mathfrak g}$
such that it is $r$-invariant, i.e. $r{\mathfrak h}\subseteq{\mathfrak
h}$. The restriction of $r$ to ${\mathfrak h}$, $r\vert_{\mathfrak h}$
is an $r$-matrix in ${\mathfrak h}$.  Since ${\mathfrak h}$ is simple,
its Killing form coincides (up to a constant factor) with the
restriction to ${\mathfrak h}$ of the Killing form in $\mathfrak g$.
Let $H\subset G$ be the subgroup with Lie algebra ${\mathfrak
h}$. Then, for $g\in H$,\ $X,Y\in {\mathfrak h}$,
\begin{eqnarray} \label{inducedPoisson}
P_{g}^{r\vert_{\mathfrak h}}(X,Y)={\frac{1}{2}}\
\tr(Xr\vert_{\mathfrak h}Y-X\Ad_{g}r\vert_{{\mathfrak h}}\Ad_{g}^{-1}Y)
\end{eqnarray}
defines a Poisson-Lie structure on $H$.

The nice point is that we can realize $H^*$, the dual Poisson-Lie
group of $H$, as a subgroup of $G_r$.
$H^*$ is simply identified with the subgroup $H_r\subset G_r$
corresponding to the Lie subalgebra $({\mathfrak h},[\cdot,\cdot]_r)$ of
$({\mathfrak g},[\cdot,\cdot]_r)$. We claim that the Poisson-Lie sigma model
with target $G$ and brane $H$ is dual to the Poisson-Lie sigma model
with target $G_r$ and brane $H_r$. That is to say, there is a
bulk-boundary duality between ${\cal P}(G,H)$ and ${\cal P}(G_r,H_r)$.

Before describing the duality we shall make some general
considerations about the properties of these branes.  The results on
the boundary conditions of the fields mentioned at the end of Section
4 are written in terms of $\psi \in \Gamma(T^*\Sigma \otimes
X^*T^*M)$.  Let us rewrite them in terms of the field $A_t$ appearing
in the action (\ref{PSgroup}).
Here the subscript $t$ refers to contraction of $A$ with vectors
tangent to $\partial\Sigma$.

When varying (\ref{PSgroup}) with respect to $g$, a boundary term
$-\int_\Sigma\dd\tr(\dd gg^{-1}\wedge A)$ appears. Its cancellation
requires $A_t \in {\mathfrak h}^\perp$ ($\perp$ means orthogonal
with respect to $\tr(\ )$). On the other hand, the continuity of
(\ref{eomG}) at the boundary imposes $P^{\sharp}A_t \in
{\mathfrak h}$. Consequently, the boundary condition for $A_t$ is
\begin{equation} \label{BCA}
A_t(\sigma) \in {\mathfrak h}^\perp\cap P_{g(\sigma)}^{\sharp
-1}{\mathfrak h},\ \sigma \in \partial\Sigma
\end{equation}
and the gauge transformation parameter $\beta$ at the boundary is
restricted by the same condition.

Condition (\ref{PoiDir}) for Poisson-Dirac branes, applied to our present 
situation reads
$${\mathfrak h}\cap P_g^{\sharp}{\mathfrak h}^\perp=0,\
\forall g\in H.$$
In particular, if $H$ is Poisson-Dirac the gauge transformations do
not act on $g$ at the boundary.

We have that $H$ is coisotropic if
$$P_g^{\sharp}{\mathfrak h}^\perp \subseteq {\mathfrak h},\ \forall g\in H$$


We show now that an $r$-invariant, simple, subgroup $H$ is a Poisson-Dirac
submanifold of $G$, and its dual $H_r$ is also Poisson-Dirac in
$G_r$. Denote by ${\mathfrak h}^\perp \subset {\mathfrak g}$ the
subspace orthogonal to ${\mathfrak h}$ with respect to ${\rm
tr}(\ )$. Firstly, the $r$-invariance of ${\mathfrak h}$ implies that
$P_g^{r\sharp}{\mathfrak h}\subseteq{\mathfrak h},\forall g\in
H$. Using that $P^{r\sharp}$ is antisymmetric we obtain that
$P_g^{r\sharp}{\mathfrak h}^\perp\subseteq{\mathfrak h}^\perp,\forall
g\in H$. Finally, recalling that ${\mathfrak h}$ simple
$\Rightarrow{{\mathfrak h}}\cap{\mathfrak h}^\perp=0$, one immediately
deduces that $H$ is Poisson-Dirac. 

Observing that ${\mathfrak h}_r$ (i.e. ${\mathfrak h}$ equipped with
the Lie bracket $[\cdot,\cdot]_{r\vert_{\mathfrak h}}$) is the same as
${\mathfrak h}$ as a vector subspace of ${\mathfrak g}$ and reasoning
as above one shows that $H_r$ is a Poisson-Dirac submanifold in $G_r$.
\vskip 2mm

Finally $H$ and $H_r$ inherit a (smooth) Poisson structure (the Dirac
bracket) from $G$ and $G_r$ respectively. They coincide with the
Poisson structures defined by $r\vert_{\mathfrak h}$ on $H$ and $H_r$
(formula (\ref{inducedPoisson}) for $H$, and analogously for $H_r$),
making them into a pair of dual Poisson-Lie groups. Notice, however,
that this induced Poisson structure does not make $H$ (resp.  $H_r$)
into a Poisson submanifold of $G$ (resp. $G_r$), so that in general it
is not a Poisson-Lie subgroup, it is so if and only if it is
coisotropic.

We now address the issue of the duality of the models with a pair of
branes $H$ and $H_r$ as above. The general picture is as follows.  In
the case of the model over $G$ with brane $H$ the space of solutions,
once reduced by the gauge transformations in the bulk, can be
identified with the universal covering of $G_rH_d\cap H_dG_r\subset
D_0$.  The symplectic form $\Omega$ (see (\ref{HeisenbergDouble})) in
$D_0$ (corresponding to free boundary conditions) becomes degenerate
in $G_rH_d\cap H_dG_r$. This reflects the existence of gauge
transformations at the boundary. Since ${\mathfrak h}$ is
$r$-invariant, there is a natural choice of gauge fixing for these
transformations: $H_rH_d\cap H_dH_r$. The pullback of $\Omega$ to
$H_rH_d\cap H_dH_r$ is nondegenerate and the infinitesimal gauge
transformations span a complementary subspace to $T_p (H_rH_d\cap
H_dH_r)$ in $T_p(G_rH_d\cap H_dG_r)$ for every $p\in H_rH_d\cap
H_dH_r$.

The dual model over $G_r$ with brane $H_r$ behaves in an analogous
way. The space of solutions of the equations of motion can be
identified with $G_dH_r\cap H_rG_d$, but there are still gauge
transformations acting on this space. The gauge fixing is given again
by considering the restriction to $H_rH_d\cap H_dH_r$, which makes
this model equivalent by bulk-boundary duality to the previous one.

The considerations of the previous paragraphs did not care about the
existence of singular points or whether the gauge fixing is local or
global.  These subtleties may depend on the concrete model. Let us
work out an example in which all properties of regularity and global
gauge fixing are met.

Take $G=SL(n,{\mathbb C})$ with the Poisson-Lie structure
(\ref{PoissonLie}) given by the standard $r$-matrix (\ref{standardr})
and
\begin{equation}
H=\left\{{\begin{pmatrix}A&0 \cr 0&I \end{pmatrix}}
\in SL(n,{\mathbb C}),\quad {\rm s.t.}\quad
\ A\in SL({k},{\mathbb
C})\right\}
\end{equation}
for a given ${k}<n$.
The dual group $H_r\subset G_r\subset G\times G$ is easily described:
\begin{equation}
H_r=\left\{(g_+,g_-)\in G_r, 
\quad {\rm s.t.}\quad
g_\pm={\begin{pmatrix}A_\pm&0 \cr 0&I \end{pmatrix}},\ A_\pm \in 
SL({k},{\mathbb
C})\right\}
\end{equation} 

In this case,
\begin{equation}
{\mathfrak h}^\perp=\left\{
\begin{pmatrix}
\lambda I&B\\
C&X
\end{pmatrix}
\in {\mathfrak sl}(n,{\mathbb C})
\right\}
\end{equation}

An easy calculation shows that $P_g^{r\sharp} {{\mathfrak h}}^\perp=0$,
i.e. $H\subset G$ is Poisson-Dirac and coisotropic. In particular, the
inclusion map $i:H\rightarrow G$ is a Poisson map and $H$ is a
Poisson-Lie subgroup of $G$.

The situation is different for the dual model with target $G_r$ and brane
$H_r$. Recall that ${\mathfrak h}_r$ is the same as ${\mathfrak h}$ as
a vector space, and hence also their orthogonal complements. In this
case ${\mathfrak h}^\perp\cap P_{(g_+,g_-)}^{r\sharp -1}{\mathfrak
h}\neq {{\mathfrak h}}^\perp$ and $H_r\subset G_r$ is not
coisotropic. In fact, in generic points
\begin{equation}
{\mathfrak h}^\perp\cap
P_{(g_+,g_-)}^{r\sharp -1}{\mathfrak h}=\left\{
\begin{pmatrix}
\lambda I&0\\
0&X
\end{pmatrix}
\in {\mathfrak sl}(n,{\mathbb C})\right\}
\end{equation}
so that the brane $H_r$ is classically admissible in the sense of
Section 4.

The solutions of the equations of motion for $g$ in the model with
target $G$ are given by
\begin{equation}
g(\sigma)={}^{(h_+(\sigma),h_-(\sigma))}{\hat g}
\end{equation}
where the action on ${\hat g}$ is given by dressing transformations
and $g(t,0),g(t,\pi)\in H$. One can always take
$(h_+(t,0),h_-(t,0))=(e,e)$ and $(h_+(t,\pi),h_-(t,\pi))\in
{H}_r$. One can fix the gauge freedom for $A_t$ at the boundary
imposing $A_t=0$. Then, $h_\pm$ are constant at every connected
component of the boundary of $\Sigma$.  Therefore, the reduced phase
space ${\cal P}(G,H)$ covers the set of pairs $((h_+,h_-),{\hat g})$
where ${\hat g}\in H$ and $(h_+,h_-)\in H_r$ and $(h_+{\hat g},
h_-{\hat g})\in H_rH_d\cap H_dH_r$.




For the Poisson-Lie sigma model with target $G_r$ the solutions of the
equations of motion for $(g_+,g_-)$ are
\begin{equation}
(g_+(\sigma),g_-(\sigma))={}^{h(\sigma)}({\hat g}_+,{\hat g}_-)\qquad
\end{equation}
with $(g_+(t,0),g_-(t,0)),(g_+(t,\pi),g_-(t,\pi))\in H_r$.

An analogous argument shows that ${\cal P}(G_r,H_r)$ is (a covering of)
the set of
pairs $(h,({\hat g}_+,{\hat g}_-))$ where $({\hat g}_+,{\hat
g}_-)\in H_r$, $h\in H$ and 
$(h{\hat g}_+,h{\hat g}_-)\in H_dH_r\cap H_rH_d$.

The duality then exchanges degrees of freedom at the boundary with
degrees of freedom in the bulk, exactly in the same way as it does for
the free boundary conditions.

Notice that the duality described so far exists only for very special
branes given by $r$-invariant subalgebras. If one considers more
general situations the result is not that clean and one has, in the dual model,
non-local boundary conditions that relate the fields at both connected
components of $\partial\Sigma$.  A more comprehensive treatment of
this case will be done elsewhere.

\section{Conclusions}

We have extended the study of Poisson-Lie sigma models to the case of
triangular $r$-matrices. To that end, we have presented a unified
treatment of the factorizable and triangular case, that allows a
convenient description of the double and the Poisson structure for the
dual model in the triangular case.

We have shown that in the triangular case the dual model is
equivalent, on-shell, to the BF-theory. This is reminiscent of what we
have in the case of factorizable $r$-matrices (see \cite{FalGaw02})
where the Poisson-Lie sigma model is equivalent to the $G/G$
Wess-Zumino-Witten model.

An aspect that is not covered in the present paper is the relation of
these models with the WZ-Poisson sigma model (\cite{KliStr}). There
are some results in the literature for the dual of a factorizable
Poisson-Lie group in connection with the $G/G$ WZW model
(\cite{SevWei},\cite{KotSchStr}). The triangular case and the twisted
version of the direct models have not been treated. We plan to turn to
them in the future.

In a previous paper \cite{CalFalGar} the relation between the
Poisson-Lie sigma models for pairs of dual groups was analysed
uncovering a duality between the gauge degrees of freedom at the
boundary in one model and those of the target field in the bulk in the
other. The result in \cite{CalFalGar} was derived using free boundary
conditions for both models. In a later paper \cite{BonZab} the
question of extending the duality to more general coisotropic branes
was addressed without conclusive results. In this paper we have given
an answer to this question by characterizing a family of branes that
can be dualized, i.e. the reduced phase space of one of the models can
be mapped by a one-to-one bulk-boundary transformation into that of
the dual model with a dual brane. The branes for which we have a dual
brane are the subgroups whose Lie algebra is $r$-invariant.

Our branes do not need to be coisotropic and moreover, if we start
with a coisotropic $r$-invariant subgroup the dual brane is an
$r$-invariant subgroup not necessarily coisotropic, as we have shown
explicitly in an example. The fact that by duality one can transform
coisotropic branes into non-coisotropic ones is a strong motivation
for the consideration and study of non-coisotropic submanifolds as
boundary conditions for the Poisson-Sigma model.

Finally, for more general branes (i.e. beyond $r$-invariant subgroups)
the situation is more complicated. It seems that for an arbitrary
brane one obtains by duality non-local boundary conditions that relate
the fields at both components of the boundary of the strip. Whether
this can be interpreted as a kind of twisted periodic boundary
conditions in the closed geometry will be the subject of further
research.


\vskip 0.5 cm

\noindent{\bf Acknowledgements:} We are indebted to Alberto Elduque
for his help with some algebraic questions at various stages of this
work. We also thank Krzysztof Gaw\c{e}dzki for useful remarks related
to the Poisson-Lie sigma model in the triangular case.

\end{document}